\documentclass[prl,twocolumn,showpacs,preprintnumbers,amsmath,amssymb]{revtex4}
\usepackage{dcolumn}
\usepackage{bm}
\usepackage{graphicx}
\usepackage{bm}
\begin{document}
\title{Excess specific heat and evidence of zero point entropy in magnetic glassy state of half-doped manganites} 

\author{A. Banerjee, R. Rawat, K. Mukherjee and P. Chaddah}
\affiliation{UGC-DAE Consortium for Scientific Research,University Campus, Khandwa Road,
Indore-452001, M.P, India.}

\begin{abstract}
We show that specific heat C$_p$ has non-Debye behavior for glassy states in half-doped manganites. Irrespective of the magnetic order or electronic states, these magnetic glasses have higher C$_p$ compared to their equilibrium counterparts. The excess C$_p$ contributed by the glassy state varies linearly with temperature similar to conventional glasses indicating tunneling in the two-level systems. These glassy states show signature of zero point entropy. Magnetic glasses can be produced simply by different field cooling protocols and may be considered ideal magnetic counterpart of the conventional glass.    
\end{abstract}
\maketitle
The nature of glass transition remains one of the most challenging problem in physics \cite{kur,and,pab,ste}. A conventional glass (CG) is viewed as a liquid with `time held still', where high temperature (T) phase is preserved at low-T in a non-ergodic state \cite{braw}. Vitrification is considered as an outcome of a dynamic singularity upon deep supercooling below glass transition temperature, T$_g$ \cite{deb}. Formation of glass depends on the efficiency with which kinetics of the first-order transformation process is arrested by crossing T$_g$, avoiding crystallization at the supercooling spinodal at T$^*$. This is evident in the orders of magnitude difference in the required cooling rates for system ranging from metallic glass to glass formers like ortho-terphenyl \cite{braw}. Such wide variation is justified on the basis of relative values of T$_g$ vis-$\grave{a}$-vis T$^*$ and brings in the important role of other control variables which can tune these characteristic temperatures \cite{gre}. This tunablity is represented by lines of the characteristic temperatures (T$^*$ and T$_g$) in the pressure (P) vs. T phase diagram and led to achieve a significant milestone when monatomic Ge could be vitrified under high pressure \cite{bhat}. In this context, recently discovered glass-like arrested state established in variety of magnetic systems and termed as `magnetic glass' opens up new possibilities \cite{cha2, mkc, kranti, wu, ban, roy, cha3, cha5, macia, sha}. Akin to the CG, magnetic glass form on crossing the T$_g$ line without completing the first-order transformation process at T$^*$, however, the second control variable used for this case is the magnetic field (H). In this scenario, for CG it is the high-T phase that is arrested below T$_g$ avoiding crystallization, whereas in magnetic glass high-T magnetic phase shows non-thermodynamic behavior below certain temperature but is completely different from spin-glass \cite{myd}, re-entrant spin-glass or cluster glass  \cite{roy1}. In the real systems, magnetic first-order transitions are broadened by disorder, naturally making it possible to cause fractional transformation of the high-T phase upon cooling and the remaining fraction falls out of equilibrium but persists down to lowest temperature as glass-like arrested state having contrasting magnetic order \cite{kranti,wu,cha5}. The coexisting phase fractions of the glass-like phase can be tuned by the cooling H, which persist even when the H is isothermally changed to zero at the lowest-T and is shown to have serious consequences on both physics and functionality \cite{cha2}.

The term magnetic glass describes a spatially ordered state where a first order magnetic transition is inhibited by lack of kinetics, and one obtains a glass-like arrested state. This arrested state is the high-T phase that exists at a low-T where a state with competing order has a lower free energy. Being the high-T equilibrium state, this arrested state has higher entropy and must be `disordered' as compared to the equilibrium state. The disorder causing this higher entropy is not always apparent, reminding us of solid $^3$He, of solid $^4$He, and of their mixtures. In half-doped manganites like Pr$_{0.5}$Sr$_{0.5}$MnO$_3$  (PSMO), the low-T state is an antiferromagnetic-insulating (AFI), whereas the arrested state is ferromagnetic-metallic (FMM) \cite{ban, cha5}. The disorder in the arrested state can be understood as that of the charge liquid. For some substitutions on B-site as in Pr$_{0.5}$Ca$_{0.5}$Mn$_{0.975}$Al$_{0.025}$O$_3$ (PCMAO), the low-T state becomes FM-M and the arrested state is AF-I. Understanding the `disorder' in these magnetic glasses is not straightforward and would also probably involve the same detailed understanding of various contributions to entropy as was necessary to understand the melting of solid heliums (albeit in a restricted pressure window) on cooling. The glass-like phase fraction in these manganites around half doping shows devitrification on heating and recrystallization to low-T equilibrium phase upon cooling after annealing \cite{cha2, cha3}, similar to glass ceramics \cite{gre}. \textit{Here we show from the calorimetric study that the glass-like state mimics the specific heat behavior observed in structural glass and has excess entropy.} Significantly, similar to the orientational glass (OG), the magnetic glasses also show linear contribution to specific heat (C$_p$) characteristic of the tunneling in two level systems (TLS) though both have translational periodicity \cite{ramos, tal}. Moreover, the advantage of the magnetic glass is the tunablity of glass-like phase fraction at the same T and H, which can be achieved with relative ease, only by cooling in different H but at the same cooling rate \cite{cha5}.  

To make an ideal trial ground from the thermodynamic viewpoint, we study two systems with contrasting ground state, PCMAO and PSMO. Consequently, their glass-like states are also contrasting, both in terms of magnetic order and electrical conductivity, which is significant since each has a characteristic contribution to the measured C$_p$.  Though the ground state of PCMAO is FMM, it is masked by hindered kinetics of the first order transformation process when cooled in H=0 and exist in glass-like AFI state \cite{ban, cha5}. Cooling in higher H, increasing amount of equilibrium FMM phase can be produced which coexist with remaining non-equilibrium AFI phase fraction at low-T when the H is reduced to zero. Contrary to this, zero field cooled (ZFC) state of PSMO is mostly AFI and it is in equilibrium at low-T \cite{cha5}. However, increasing amount of glass like FMM phase fraction can be accrued by cooling in higher H and substantial fraction of them remain even when the H is reduced to zero at low-T and coexist with the remaining equilibrium AFI phase fraction. 

The PCMAO and PSMO samples are the same used in Ref.\cite{ban,cha5}. The C$_p$ measurements are carried out using semi-adiabatic heat pulse \cite{idas,akr} technique with absolute accuracy $\approx$ 0.5\% in an 8 Tesla cryostat. For this study, C$_p$ is measured while warming after cooling the samples in different fields. For PCMAO, C$_p$ as a function of T is measured while warming in H=0 after cooling the sample in 0, 5 and 8 Tesla and are denoted as (0, 0), (5, 0) and (8, 0) respectively. The ZFC state of PCMAO is glassy AFI, whereas, 5 and 8T cooled states have respectively about 65\% and 89\% equilibrium FMM phase fraction at low-T in H=0 coexisting with the remaining glassy AFI phase fractions. For PSMO, the C$_p$ vs. T is measured while warming in H=0, after cooling the sample in 0 and 5 Tesla, and are denoted as (0, 0) and (5, 0) respectively. Along with these, C$_p$ vs. T is measured while warming PSMO in 5 Tesla after cooling in same H denoted as (5, 5) state.  For PSMO, the ZFC state is also predominantly AFI but contrary to PCMAO, it is in equilibrium. The 5T cooled states have large fraction of glass-like FMM phase at low-T which coexist with the remaining equilibrium AFI phase fractions. When the field is isothermally reduced to zero at low-T, a fraction of glass-like FMM (accrued while cooling in 5T) converts to equilibrium AFI phase. However, a significant fraction, much more than the ZFC state, persist in glass-like FMM phase and coexist with remaining equilibrium AFI phase. This is justified from the heuristic phase diagram made from phenomenology \cite{cha5}. Thus, C$_p$ is measured in two samples after creating different fractions of glassy phases having contrasting magnetic long-range order with structural periodicity.
\begin{figure}[htbp]
	\centering
		\includegraphics{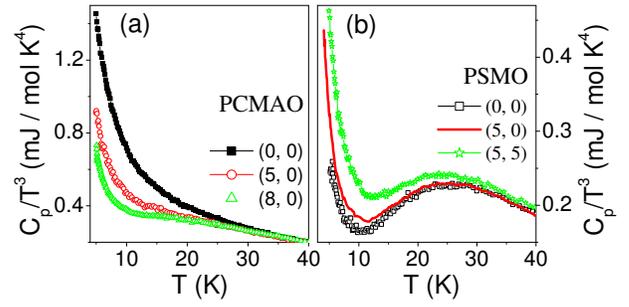}
	\caption{Low-temperature specific heat C$_p$/T$^3$ vs. T under different experimental protocols (mentioned in the text). (a) for the PCMAO sample (b) for PSMO sample.}
	\label{fig:Fig1}
\end{figure}

The measured C$_p$ of PCMAO and PSMO clearly show strong deviation from the expected Debye T$^3$ dependence at low-T when C$_p$/T$^3$ is plotted as a function of T in Fig. 1(a) and 1(b) respectively. C$_p$/T$^3$ vs. T show qualitatively similar signatures to that of structural glass or the OG which has long range structural order \cite{ramos, tal}. At low-T a clear upturn is seen which is an unambiguous indication of the presence of glassy phase. Such upturn is related to the presence of TLS excitation and is in contrast to the typical Debye behavior observed in crystal (which does not shows an upturn at low temperatures). It is rather remarkable that, C$_p$ shows progressive increase at the same T as the glassy phase fraction is increased by different field cooling protocols in both the samples. The glass-like AFI phase fraction in PCMAO varies in states as (0, 0) $>$ (5, 0) $>$ (8, 0). On the contrary, in PSMO the glass-like FMM phase fraction varies as (0, 0) $<$ (5, 0) $<$ (5, 5). 
To confirm the presence of glass-like phase fractions in both the samples C$_p$/T vs T$^2$ plots are given in Figs. 2(a) and 2(b). All the plots exhibit good linear fit with positive intercepts on y-axis at T=0, indicating that C$_p$ is essentially made of linear and cubic terms. The data in Fig. 2 are fitted with an expression of the form
\begin{center}
C$_p$ = $\gamma$T +$\beta$T$^3$............(1) 
\end{center}
\begin{figure}[htbp]
	\centering
		\includegraphics{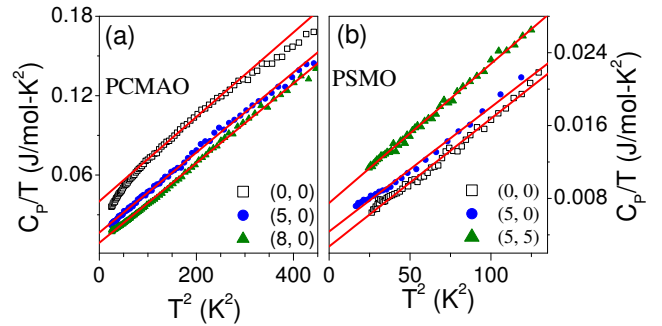}
	\caption{Specific heat (C$_p$) plotted as C/T vs T$^2$ under different experimental protocols (mentioned in the text). (a) for PCMAO sample and (b) for PSMO sample. The solid lines are fits to equation (1).}
	\label{fig:Fig2}
\end{figure}
where $\gamma$ and $\beta$ are fitting constants. In the fitting equation a T$^5$ term, which is a standard correction in the lattice contribution is not added as satisfactory fits were obtained using equation (1) and no other additional correction term improve the fitting significantly with physically significant constants.  The values of $\gamma$ and $\beta$ for both PCMAO and PSMO, each prepared in states with  three different fractions of glassy phases by different field cooling protocols are given in the Table 1. Same table also contains the corresponding Debye temperatures ($\theta$$_D$) obtained from the standard expression $\theta$$_D$ = (12$\pi$$^4$pR/5$\beta$)$^{1/3}$, where R is the ideal gas constant and p is the number of atoms per formula units. The values of $\gamma$ and $\beta$ for both the samples are in general agreements with those observed in manganites \cite{har,ham}.

\begin{table}
\caption{\label{tab:table 1}Summary of the fitting results for the C$_p$ vs temperature data shown in Fig. 1 and inset of Fig 1 for PCMAO and PSMO samples respectively. The definition of the co-efficient are given in the text}
\begin{ruledtabular}
\begin{tabular}{cccc}
\textit{Sample} &&  \textbf{PCMAO}\\
\hline
Protocols &(0T, 0T) &(5T, 0T) &(8T, 0T)\\
\hline
$\gamma$ (mJ/molK$^2$) &40.7 &16.5  &8.7 \\ 
\hline
$\beta$(mJ/molK$^4$) &0.319  &0.302  &0.301 \\ 
\hline
$\theta$$_D$ (K) &294 &299  &300 \\
\hline
\hline
\textit{Sample} &&  \textbf{PSMO}\\
\hline
Protocols &(0T, 0T) &(5T, 0T) &(5T, 5T)\\
\hline
$\gamma$ (mJ/molK$^2$) &3.18 &4.6 &7.5 \\ 
\hline
$\beta$(mJ/molK$^4$) &0.132  &0.135 &0.153  \\ 
\hline
$\theta$$_D$ (K) &393 &391 &375\\
\end{tabular}
\end{ruledtabular}
\end{table}

However, unlike metallic systems, the linear term does not arise entirely from the standard free carriers contribution. This is clear from the $\gamma$ values of PCMAO given in Table 1. In PCMAO, the FMM phase fractions increase with the cooling H i.e. the metallic phase fraction progressively increases from (0,0) $<$ (5,0) $<$ (8,0) states whereas  $\gamma$ values decreases from 40.7 to 16.5 to 7.9 mJ/molK$^2$ respectively. However, neither $\beta$ nor $\theta$$_D$ show similar drastic change in this same sample. Moreover, the value of $\gamma$ in (0, 0) state (AFI) for PCMAO sample is a bit high. Such linear term found in insulating manganites is generally related to disorder affecting the spin, charge and orbital degrees of freedoms. Another manifestation of disorder is the appearance of downward curvature below 9K (Fig. 2a). This is not uncommon and has already been observed in half doped manganites \cite{har}. This is attributed to the fact that the charge-ordered state in Pr$_{0.5}$Ca$_{0.5}$MnO$_3$ is destabilized without being completely destroyed by 2.5\% Al substitution \cite{sunil}. In PCMAO sample cooling in a high field destroys the AFM phase \cite{ban} and hence there is no sudden increase in C$_p$, leading to a linear C$_p$/T vs. T$^2$ behavior. Hence, two half-doped manganites (with contrasting first-order transition) shows most genuine fingerprints of glassy properties and the behavior of this phase is similar to that of CG.  

We reconfirm that the signature of the glassy phase is devoid of any artifact arising because of the two structures associated with distinct magnetic orders in the coexisting phases. To isolate the contribution to C$_p$ arising from glassy feature, the difference between the specific heats ($\Delta$C$_p$) of the phases consisting of maximum glassy fraction to those of lesser fractions of glass are plotted in Fig. 3 as a function of T.  Hence, for PCMAO,  the difference between C$_p$s of (0,0) - (5,0) as well as (0,0) - (8,0) are plotted in Fig. 3(a). Similarly, for PSMO the difference between C$_p$s of (5,5) - (5,0) and (5,5) - (0,0) states are plotted as function of T Fig. 3(b). It is rather significant that for both the samples  $\Delta$C$_p$ varies linearly with T. Moreover, the excess specific heat,  $\Delta$C$_p$, is directly related to the fraction of the glassy phase but does not arise from the linear contribution of the electronic C$_p$, as it is not related to conductivity behavior which is opposite in these two samples. Further,  $\Delta$C$_p$ does not fit to equation (1) with physically acceptable coefficient for T$^3$ term ($\beta$), which imply that the excess specific heat has insignificant contribution from the underlying structure and arise from the non-Debye like contribution of the magnetic glassy state.
\begin{figure}[htbp]
	\centering
		\includegraphics{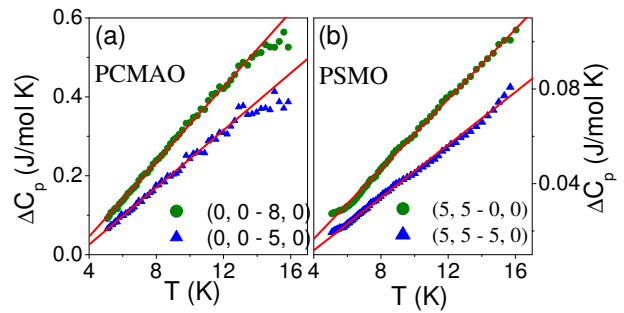}
	\caption{Difference in Specific heat (see text) plotted as a function of T (a) for PCMAO sample and (b) for PSMO sample. Solid lines are linear fit to the data.}
	\label{fig:Fig3}
\end{figure}

Thus there is clear indication that the magnetic glasses are like CG as linear C$_p$ vs. T is a universal feature of glassy systems at low-T \cite {zel}. This also signifies that like CG or OG, magnetic glass also has degenerate states and C$_p$ shows signature of tunneling between such states \cite{ande}. Hence, it appears that the magnetic glass possess tunneling states or is a TLS and are similar to CG. It may be noted that the signature of TLS in CG appears at very low-T ($\leq$ 1K) and is related to the energy scale of the atomic motion within the degenerate states \cite{zel,ande}. Whereas, in OG linear C$_p$ is observed up to relatively higher-T ($~$a few Kelvin) because of the higher energy scales involved in rotational motion of the molecules \cite{ramos,tal}. We venture to suggest that the observation of linear C$_p$ vs. T behavior in these magnetic glasses up to fairly higher-T can be attributed to the higher energy scales of their degenerate magnetic states. In these manganites around half doping, this energy scale is possibly related to the barrier between the AFI and FMM states. Magnetic fields of the order of a few Tesla are required to cause the field-induced transition and this energy scale is probably linked with the TLS type behavior resulting in observation of linear C$_p$ up to significantly higher-T. It is important to note that evidence of TLS type behavior with similar energy scale in other manganite was given by Raychaudhuri et al.\cite{akr}.

Now we calculate the entropy of the glassy phase fraction (S$_G$) from the excess specific heat ( $\Delta$C$_p$), integrating  $\Delta$C$_p$/T. In Fig. 4 we plot S$_G$ vs. T only for the maximum $\Delta$C$_p$ case. Assumes Ising like states, entropy at high-T should extrapolate to S = Rln2 when all possible states are explored \cite{lau}. In this context, it is to be noted that the difference between the integrated entropy and Rln2 is known as zero point entropy \cite{ke}. In our case, the value of entropy as observed from Fig. 4 is less than the ideal value because all the states in the system are not available. This gives a clear indication of the presence of zero point entropy for these systems and specific evidence about the presence of the glassy phase. The situation develops in-order to avoid the entropy crisis (Kauzmann paradox) which may arise while cooling, if the entropy of a supercooled liquid become smaller than that of a crystal. It results in the formation of a glassy state with higher entropy than that of a crystalline solid.
\begin{figure}[htbp]
	\centering
		\includegraphics{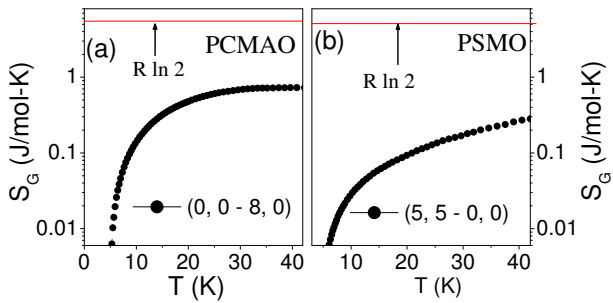}
	\caption{Entropy of the glassy phase plotted as a function of T (a) for PCMAO and (b) for PSMO sample}
	\label{fig:Fig4}
\end{figure}

In summary, we show from calorimetry the presence of glass-like states but in two manganites around half-doping with contrasting ground states. In both the systems, C$_p$ increases with the glass-like phase fraction in spite of their contrasting order. For the same sample, the difference in C$_p$ between states created with largest fraction of glass to the smaller fractions of glass is attributed as the excess specific heat ($\Delta$C$_p$) arising from glassy phase. Similar to other glasses,  $\Delta$C$_p$ increases linearly with T and is considered to be related to the tunneling in a TLS. The calculated contribution to entropy of these magnetic glassy states show signatures of zero point entropy. Thus it will be interesting to look for other universal features observed in glasses like boson peak observed in inelastic scattering of neutrons or photons\cite{boson}.

We thank Kranti Kumar, A. K. Pramanik and Pallavi Kushwaha for their help in the experimental work.

\newpage


\begin{thebibliography}{}
\bibitem[1]{kur} J. Kurchan, Nature (London) \textbf {433}, 222 (2005).
\bibitem[2]{and} P. W. Anderson, Science \textbf {267}, 1615 (1995).
\bibitem[3]{pab} P. G. Debenedetti and F. H. Stillinger, Nature (London) \textbf {410}, 259 (2001).
\bibitem[4]{ste} S. A. Kivelson and G. Tarjus, Nature Mater. \textbf {7}, 831 (2008).
\bibitem[5]{braw} S. Brawer, \textit{Relaxation in viscous liquid and glasses} The American ceramic society, Inc., Columbus, Ohio (1985).
\bibitem[6]{deb} Pablo G. Debenedetti, \textit{Metastable Liquids concept and principle} Princeton University Press, Princeton(1995).
\bibitem[7]{gre} A. L. Greer, Science \textbf {267}, 1947 (1995).
\bibitem[8]{bhat} M. H. Bhat \textit{et al}., Nature \textbf {448}, 787 (2007).
\bibitem[9]{cha2} P. Chaddah \textit{et al}., Phys. Rev. B \textbf {67}, 100402(R) (2008).
\bibitem[10]{mkc} M. K. Chattopadhyay \textit{et al}., Phys. Rev. B \textbf{72}, 180401 (R) (2005).
\bibitem[11]{kranti} Kranti Kumar \textit{et al}., Phys. Rev. B \textbf{73}, 184435 (2006).
\bibitem[12] {wu} W. Wu \textit{et al}., Nature Mater \textbf{5}, 881 (2006).
\bibitem[13] {ban} A. Banerjee \textit{et al}., Phys. Rev. B \textbf{74}, 224445 (2006).
\bibitem[14] {roy} S. B. Roy \textit{et al}., Phys. Rev. B 75, 184410 (2007). 
\bibitem[15]{cha3} A. Banerjee \textit{et al}., J. Phys.: Condens. Matter \textbf {20} 255245 (2008); \textit{ibid} \textbf {21} 026002 (2009) and references therein. 
\bibitem[16]{cha5} A Banerjee \textit{et al}., J. Phys.: Condens. Matter \textbf {18} L605 (2006).
\bibitem[17]{macia} F. Maci$\grave{a}$ \textit{et al}., Phys. Rev. B \textbf {77}, 012403 (2008). 
\bibitem[18]{sha} P. A. Sharma \textit{et al}., Phys. Rev. B \textbf {78}, 134205 (2008). 
\bibitem[19]{myd} J. A. Mydosh, \textit{Spin glasses} (Taylor and Francis, London, 1992). 
\bibitem[20] {roy1} S. B. Roy and M. K. Chattopadhyay, Phys. Rev. B \textbf {79}, 052407 (2009). 
\bibitem[21]{ramos} M. A. Ramos \textit{et al}., Phys. Rev. Lett. \textbf {78}, 82 (1997); C. Talon et al., Phys. Rev. B \textbf {58}; 745 (1998), C. Talon \textit{et al}., Phys. Rev. B  \textbf {65}, 012203 (2001).
\bibitem[22]{tal} C. Talon \textit{et al}., Phys. Rev. B  \textbf {66}, 012201 (2002). 
\bibitem[23]{idas} R. Rawat and I. Das, Phys. Rev. B  \textbf {64}, 052407 (2001). 
\bibitem[24]{akr} A. K. Raychaudhuri \textit{et al}., Phys. Rev. B  \textbf {64}, 165111 (2001). 
\bibitem[25]{har} V. Hardy \textit{et al}., Phys. Rev. B \textbf {67}, 024401 (2003).
\bibitem[26]{ham} J. J. Hamilton \textit{et al}., Phys. Rev. B  \textbf {54}, 14926 (1996); M. R. Lees et al., Phys. Rev. B  \textbf {59}, 1298 (1999); V. N. Smolyaninova \textit{et al}., Phys. Rev. B  \textbf {58}, 14725 (R) (1998). 
\bibitem[27]{sunil}Sunil Nair and A. Banerjee Phys. Rev. Lett. \textbf {93}, 117204 (2004).
\bibitem[28]{zel} R. C. Zeller and R. O. Pohl, Phys. Rev. B \textbf {4}, 2029 (1971). 
\bibitem[29]{ande} P. W. Anderson \textit{et al}., Philos. Mag. \textbf {25}, 1 (1972).
\bibitem[30]{lau} G. C. Lau \textit{et al}., Nature Physics \textbf {2}, 249 (2006). 
\bibitem[31]{ke} X. Ke \textit{et al}., Phys. Rev. Letts. \textbf {99}, 137203 (2007).
\bibitem[32] {boson} H. Shintani and H. Tanaka, Nature Mater. \textbf{7}, 870 (2008).
\end{thebibliography}
\end{document}